\documentclass[12pt,letterpaper]{article}
\usepackage{t1enc}
\usepackage[latin1]{inputenc}
\usepackage[english]{babel}
\usepackage{graphicx}
\usepackage{amsmath}
\begin{document}

\title{Maximum Likelihood Estimation of Nonnegative Trigonometric Sum Models Using a Newton-like Algorithm on Manifolds}
\renewcommand{\baselinestretch}{1.00}
\author{$*$ Fernández-Durán, J.J. and $**$ Gregorio-Domínguez, M.M. \\
 $*$ Department of Statistics \\
 $**$ Department of Actuarial Science \\
Instituto Tecnológico Autónomo de México.\\
R\'io Hondo No. 1, Col. Progreso Tizap\'an, C.P. 01080, M\'exico D.F., M\'exico \\
e-mail: jfdez@itam.mx}
\date{}
\maketitle

\begin{abstract}
In Fernández-Durán (2004), a new family of circular distributions based
on nonnegative trigonometric sums (NNTS models) is developed. Because the parameter space of this family is the surface of the hypersphere,
an efficient Newton-like algorithm on manifolds is generated in order to obtain the maximum
likelihood estimates of the parameters.
\end{abstract}

\textbf{Keywords}: Differential Geometry, Maximum Likelihood Estimation, Newton Algorithm,  Nonnegative Fourier Series, Smooth Riemann Manifold.

\section{Introduction}

The probability density function, $f(\theta;\underline{c})$, of a circular random variable $\Theta \in (0,2\pi]$ must be nonnegative and periodic
($f(\theta + 2k\pi; \underline{c})=f(\theta; \underline{c})$) for any integer $k$ where $\underline{c}$ is the vector of parameters. Practical examples of circular random variables include the wind directions at
different monitoring stations, the directions taken by an animal, the times at which a person conducts a daily activity, the time of occurrence of different events, and many others. Based on the results of Féjer (1915), Fernández-Durán (2004) derived
a family of circular distributions based on nonnegative trigonometric sums (see also Fernández-Durán, 2007). In short,
the nonnegative trigonometric sum is expressed as a squared norm of a complex number. The circular density function based on
nonnegative trigonometric sums (NNTS density) is expressed as

\begin{equation}
f(\theta ; M, \underline{c}) = \left|\left|\sum_{k=0}^{M}c_k e^{ik\theta}\right|\right|^2=\sum_{k=0}^M\sum_{l=0}^M c_k\bar{c}_l e^{i(k-l)\theta}.
\end{equation}
Note that $i=\sqrt{-1}$ and, $c_k=c_{rk} + ic_{ck}$ are complex numbers for $k=0, \ldots, M$, and, $\bar{c}_k=c_{rk} - ic_{ck}$ is the conjugate of $c_k$. To integrate to one, it is necessary
to impose the following constraint in the $c$ parameters.
\begin{equation}
\sum_{k=0}^M ||c_k||^2 = \frac{1}{2\pi}.
\end{equation}
Note that $c_{c0}=0$ and $c_{r0} \ge 0$; i.e., $c_0$ is a nonnegative real number.
Thus, the $c$ parameter space corresponds to the surface of a $2M + 1$ dimensional
hypersphere. This family of circular distributions has the advantage of being able to fit datasets that present multimodality and/or skewness because the density function can be expressed as a mixture of
multimodal circular distributions. The total number of $c$ free parameters is equal to $2M$.

The main objective of this paper is to develop an efficient Newton-like optimization algorithm on the surface of a hypersphere that corresponds to
a Riemann manifold, in order to obtain the maximum likelihood estimates of the $c$ parameters.

The paper is divided into five sections, including this introduction. The second section presents a convenient, alternative way to express likelihood
functions for continuous and grouped data in the univariate case. Given these convenient expressions for likelihood functions, in the third section an efficient
Newton-like algorithm is developed for maximizing the log-likelihood function on the surface of the hypersphere. The proposed algorithm is a particular case
of a Newton-like algorithm for scalar functions on Stiefel manifolds (Absil \emph{et al.}, 2008, Manton, 2002, Balogh, 2004). In the fourth section, some
applications of the proposed algorithms to real continuous and grouped datasets are presented. Finally, the conclusions of the present work are presented in the fifth section.

\section{Likelihood Functions}

\subsection{Continuous Data}

Let $\theta_1, \theta_2, \ldots, \theta_n$ be a random sample of univariate circular random variables from a population with density function $f(\theta ; M, \underline{c})$, which is a member of
the NNTS family with parameters $\underline{c}$ and M. The density function of $\theta_j$ is given by
\begin{equation}
f(\theta_j ; M, \underline{c}) = \left|\left|\sum_{k=0}^{M}c_k e^{ik\theta_j}\right|\right|^2=\sum_{k=0}^M\sum_{l=0}^M c_k\bar{c}_l e^{i(k-l)\theta_j}
\end{equation}
which can be written in the following quadratic form:
\begin{equation}
f(\theta_j ; M, \underline{c}) = \underline{c}^H \underline{e}_j \underline{e}_j^H \underline{c} = \underline{c}^H E_j \underline{c}.
\end{equation}
Note that $\underline{c}=(c_0, c_1, \ldots, c_M)^T$, $\underline{e}_j=(1, e^{-i\theta_j}, e^{-2i\theta_j},e^{-3i\theta_j}, \ldots, e^{-Mi\theta_j})^T$, and
$\underline{c}^H$ indicate the Hermitian transpose of the vector $\underline{c}$ that corresponds to the transpose and conjugate of the vector of complex numbers $\underline{c}$, and $\underline{c}^T$ is the transpose of $\underline{c}$. Then, the likelihood for a random sample $\theta_1, \ldots, \theta_n$, denoted by $L(M, \underline{c} \mid \theta_1, \ldots, \theta_n)$, is calculated as
\begin{equation}
L(M,\underline{c} \mid \theta_1, \ldots, \theta_n) = \prod_{k=1}^n \underline{c}^H \underline{e}_k \underline{e}_k^H \underline{c} =
\prod_{k=1}^n \underline{c}^H E_k \underline{c}
\end{equation}
and the corresponding log-likelihood function is
\begin{equation}
l(M,\underline{c} \mid \theta_1, \ldots, \theta_n) = \sum_{k=1}^n \ln(\underline{c}^H \underline{e}_k \underline{e}_k^H \underline{c}) =
\sum_{k=1}^n \ln(\underline{c}^H E_k \underline{c}).
\end{equation}

\subsection{Grouped Data}

Let $(a_j,b_j]$ for $j=1, \ldots, Q$ be a partition of the interval $(0,2\pi]$, i.e., $(0,2\pi] = \cup_{k=1}^Q (a_j,b_j]$ and
$(a_j,b_j] \cap (a_k,b_k] = \emptyset$ for $j \ne k$, and let
$N_1, N_2, \ldots, N_Q$ be the total number of observations in each of the intervals in the partition. Let $N_k$ be the total number of observations
in the interval $(a_k,b_k]$ for $k=1, \ldots, Q$. This type of data is called grouped or incidence data. The likelihood function is
\begin{equation}
L(M,\underline{c} \mid N_1, \ldots, N_Q) = \prod_{k=1}^Q \left( F(b_k;M,\underline{c}) - F(a_k;M,\underline{c}) \right)^{N_k}
\end{equation}
where $F(b_k;M,\underline{c})$ is the accumulated distribution function of the NNTS density at $b_k$. Note that the accumulated distribution function is
obtained as
\begin{equation}
F(b_k;M,\underline{c}) = \int_0^{b_k} f(\theta;M,\underline{c})d\theta = \int_0^{b_k} \underline{c}^H E \underline{c} d\theta =
\underline{c}^H \left(\int_0^{b_k} E d\theta \right) \underline{c}
\end{equation}
where $\left(\int_0^{b_k} E d\theta \right)$ integrates each element of the matrix $E$. The elements of matrix $E$ are of the form
$e^{ir\theta}$ for $r=-M, -M+1, \ldots, 0, \ldots, M-1, M$. The value of the integral is equal to $\frac{i}{r}(1 - e^{irb_k})$ for $r \ne 0$ and
equal to $b_k$ for $r=0$. The likelihood can again be expressed as a product of quadratic forms with respect to $\underline{c}$.

\section{The Newton-like Algorithm}

Because the $c$ parameter space corresponds to the surface of the hypersphere, to obtain the maximum likelihood estimates, it is possible to apply
a Newton-like optimization algorithm on manifolds (Absil \emph{et al.}, 2008). Basically, a smooth manifold is a surface that can be approximated locally by a
hyperplane. For a point on the manifold, the approximating hyperplane is known as the tangent space. Then, a real function on a manifold can be maximized by searching for optima in the directions of movement on the tangent space and reprojecting onto the manifold. In differential geometry, the reprojection operation is called a retraction.
In this paper, the optimization problem of obtaining the maximum likelihood estimates is equivalent to maximizing a real function (i.e., the log-likelihood) on a manifold (that is, the surface of the hypersphere). The goal of the Newton-like algorithm on manifolds is to obtain the solutions of
\begin{equation}
gradl(\underline{c})=\underline{0}
\end{equation}
where $gradl(\underline{c})$ represents the gradient of the log-likelihood function $l$ at the point $\underline{c}$. The solutions to this equation correspond
to critical points of the real function $l$ on the surface of the hypersphere. The maximum likelihood estimate of $\underline{c}$ is a critical point of $l$.
The Newton method on manifolds is an iterative algorithm defined by the following steps, which are from Absil \emph{et al.} (2008):
\begin{enumerate}
\item Select an initial point $\underline{c}_0$.
\item For $k=1,2, \ldots$, solve the Newton equation
\begin{equation}
Hessl(\underline{c}_k)\underline{\eta}_k = -gradl(\underline{c}_k)
\end{equation}
for the unknown $\underline{\eta}_k$ in the tangent space at $\underline{c}_k$.
\item Set $\underline{c}_{k+1} = R_{\underline{c}_k}(\underline{\eta}_k)$ where $R_{\underline{c}_k}$ is the retraction from the tangent space onto the manifold at
$\underline{c}_k$.
\end{enumerate}
The algorithm terminates when the norm of the gradient or the norm of the difference $gradl(\underline{c}_k) - gradl(\underline{c}_{k-1})$ is less than a prespecified error. For the case considered in this paper, we use
differentiation rules of real functions of a complex vector to derive
\begin{equation}
gradl(\underline{c}) = P_{\underline{c}}\left(\frac{\partial l(\underline{c})}{\partial \underline{c}^H} \right)^H=
\left(\frac{1}{2\pi}I - \underline{c}\underline{c}^H \right)\left(
\sum_{k=1}^n \frac{\underline{c}^H\underline{e}_k\underline{e}_k^H}{\underline{c}^H\underline{e}_k\underline{e}_k^H\underline{c}} \right)^H =
\frac{1}{2\pi}\sum_{k=1}^n \frac{\underline{e}_k}{\underline{c}^H\underline{e}_k} - n\underline{c}
\end{equation}
where $P_{\underline{c}}$ is the projection onto the tangent space. For the case of the hypersphere, $P_{\underline{c}} = \frac{1}{2\pi}I - \underline{c}\underline{c}^H$ can be used.
Note that the expected value of the gradient is equal to zero, and
the Hessian matrix is obtained as
\begin{equation}
Hessl(\underline{c}) = P_{\underline{c}} \Delta gradl(\underline{c}) =
-P_{\underline{c}}\left(\sum_{k=1}^n \frac{\underline{e}_k\underline{e}_k^H}{\underline{c}^H\underline{e}_k\underline{e}_k^H\underline{c}}  \right).
\end{equation}
Fisher's information matrix, $\imath$, which corresponds to
the negative of the expected value of the Hessian, is equal to
\begin{equation}
\imath=-E\left(Hessl(\underline{c})\right)=nP_{\underline{c}}.
\end{equation}
Instead of using the Hessian in the Newton algorithm, we prefer to use Fisher's information matrix in the same way as in Fisher's method of scoring (Shao, 2003, Lange, 2004, Thisted, 1988). The modified Newton algorithm consists
of the following steps:
\begin{enumerate}
\item Select an initial point $\underline{c}_0$.
\item For $k=1,2, \ldots$, solve the Fisher's scoring equation
\begin{equation}
\imath\underline{\eta}_k = gradl(\underline{c}_k)
\end{equation}
and because
\begin{equation}
\imath\underline{\eta}_k = nP_{\underline{c}}\underline{\eta}_k =n\underline{\eta}_k
\end{equation}
and
\begin{equation}
gradl(\underline{c}_k) =
\frac{1}{2\pi}\sum_{r=1}^n \frac{\underline{e}_r}{\underline{c}_k^H\underline{e}_r} - n\underline{c}_k,
\end{equation}
the Fisher's scoring equation has the following solution for $\underline{\eta}_k$,
\begin{equation}
\underline{\eta}_k = \frac{1}{2 \pi n}\sum_{r=1}^n \frac{\underline{e}_r}{\underline{c}_k^H\underline{e}_r} -
\underline{c}_k.
\end{equation}
\item Set $\underline{c}_{k+1} = R_{\underline{c}_k}(\underline{\eta}_k)  =  \frac{\underline{\eta}_k}{\sqrt{2 \pi} ||\underline{\eta}_k||}$
\end{enumerate}
where $R_{\underline{c}_k}$ is a retraction from the tangent space onto the manifold for $\underline{c}_k$. In particular, we use
\begin{equation}
R_{\underline{c}_k}(\underline{\eta}_k) = \frac{\underline{\eta}_k + \underline{c}_k}
{\sqrt{2\pi}||\underline{\eta}_k + \underline{c}_k||}.
\end{equation}

We terminate the algorithm when the difference $||\underline{c}_{k+1} - \underline{c}_k||$ is less than a prespecified error.

The modifications required to apply the proposed algorithm to grouped data is direct because the log-likelihood function
to be maximized has the same basic form as the one we treated above.

For the practical application of the algorithm, it is possible to work with
vectors $\underline{c}$  with unit norms in order to avoid the correction terms
related to the factor $\frac{1}{2\pi}$. This is facilitated by the fact that it
is equivalent to a modified likelihood that is obtained by multiplying
the original likelihood by a constant factor. In relation to the initial point
$\underline{c}_0$, one can use a random initial point or the normalized average of the $\underline{e}_k$ statistics. In our experience, using the normalized average of the $\underline{e}_k$ statistics as an initial point has worked very well for different datasets.

\begin{figure}[t]
\center{
\includegraphics[scale=.5, bb= 40 170 530 630]{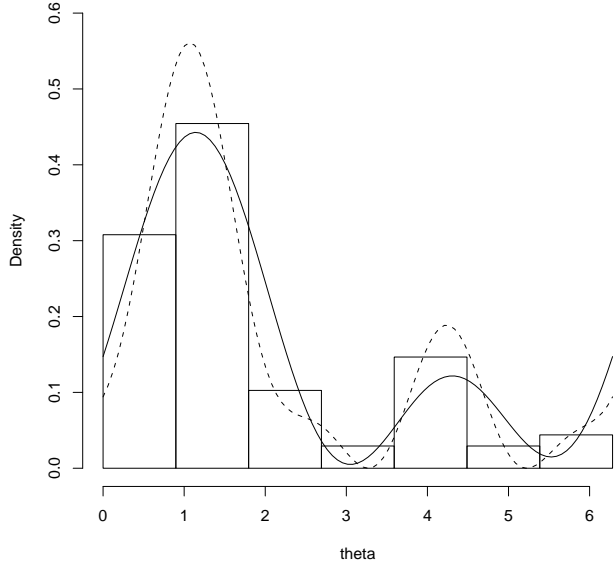}
\caption{Turtle data: Histogram of raw data and the best AIC (dashed line) and BIC (solid line) NNTS-fitted densities.
\label{Rawturtles}
}
}
\end{figure}

The proposed algorithm has been compared with results derived from sequential quadratic programming (SQP) and the Nelder-Mead optimization algorithm
in many different datasets; the proposed algorithm shows much faster convergence for many different random initial points, and in contrast to SQP and the Neder-Mead algorithm, the proposed modified Newton algorithm
usually converges to the same point. The optimality properties of Newton algorithms on manifolds
and its convergence properties are presented in Absil \emph{et.al.} (2008), Manton (2002) and Balogh (2004). Of course, as in any iterative optimization algorithm, it is important to run
the algorithm many times using different initial random points to try to find
the global maximum of the log-likelihood function.

The proposed algorithm is implemented using the statistical software $R$, particularly $CircNNTSR$ $v1.0$ (see Fernández-Durán and
Gregorio-Domínguez, 2009).

\section{Examples}

The first example refers to univariate continuous circular data and consists
of the directions taken by 76 turtles after treatment. This data set is taken from Fisher (1993, pp. 241), who
in turn took it from Stephens (1969).  The second example consists of the accumulated monthly number of deaths by suicide in England and Wales during the period 1982-1996 as an example of grouped data (Yip \emph{et al.}, 2000).

\subsection{Continuous Data}

Figure \ref{Rawturtles} presents the raw data histogram for the turtle data and the best-fitted NNTS models. Table \ref{turtlesloglikAICBIC} presents the values of the log-likelihood,
Akaike's Information Criterion (AIC), and the Bayesian Information Criterion (BIC) for
NNTS models for $M=0, 1, \ldots, 10$. This dataset was analyzed previously by Fernández-Durán (2004) using SQP to obtain the maximum likelihood estimates.
Contrary to SQP and the Nelder-Mead optimization method, the proposed Newton-like algorithm presents much faster and more stable convergence properties.

\begin{table}[h]
\begin{center}
\scalebox{0.8}{
\begin{tabular}{c||ccc}
\hline
     & \multicolumn{3}{|c}{NNTS model}   \\
$M$  & loglik ($l_{M}$) & AIC & BIC  \\
\hline
\hline
0   & -139.68  & 279.36  & 279.36   \\
1   & -122.33  & 256.66  & 261.32   \\
2   & -107.97  & 223.94  & 233.26*   \\
3   & -107.94  & 227.87  & 241.86   \\
4   & -103.96  & 223.92*  & 242.57   \\
5   & -103.33  & 226.66  & 249.97   \\
6   & -102.72  & 229.45  & 257.42   \\
7   & -102.49  & 232.98  & 265.61   \\
8   & -100.88  & 233.77  & 271.06   \\
9   & -100.50  & 237.00  & 278.95   \\
10  & -100.27  & 240.54  & 287.15   \\
\hline
\end{tabular}}
\caption{Turtle data: Log-likelihood, AIC, and BIC values for the NNTS models fitted by the proposed Newton-like algorithm on the surface of a hypersphere.
An asterisk (*) marks the best AIC and BIC models.  \label{turtlesloglikAICBIC} }
\end{center}
\end{table}

\subsection{Grouped Data}

Table \ref{Datasuicides} presents the raw monthly suicide data by sex taken from Yip \emph{et al.} (2000).
For grouped data, instead of using AIC or BIC criteria to select the best models among the considered models that use $M=0, 1, \ldots, 6$, we apply
likelihood ratio tests to compare the maximized log-likelihood values for models with $M=0,1, \ldots, 5$, $l_M$, with the maximized log-likelihood of the saturated model, $l_S$, that corresponds, in this case, to an NNTS model with $M=6$. In this strategy
for model selection, $-2(l_{M} - l_{S})$ is asymptotically distributed as a chi-squared random variable with $12-2M$ degrees of freedom.
The most parsimonious model was selected. Table \ref{suicidesloglikAICBIC} presents the values of the log-likelihood and likelihood ratio test statistics for
NNTS models with $M=0, 1, \ldots, 5$.

\begin{table}[t]
\begin{center}
\scalebox{0.8}{
\begin{tabular}{c|cccccccccccc}
\hline
 & Jan & Feb & Mar & Apr & May & Jun & Jul & Aug & Sep & Oct & Nov & Dec \\
\hline
\hline
Female & 1362 & 1244 & 1496 & 1452 & 1448 & 1376 & 1370 & 1301 & 1337 & 1351 & 1416 & 1226 \\
\hline
Male & 3755 & 3251 & 3777 & 3706 & 3717 & 3660 & 3669 & 3626 & 3481 & 3590 & 3605 & 3392 \\
\hline
\end{tabular}}
\caption{Monthly suicides in England and Wales for the period 1982-1996 (Yip \emph{et al.}, 2000).
\label{Datasuicides}}
\end{center}
\end{table}

\begin{table}[h]
\begin{center}
\scalebox{0.8}{
\begin{tabular}{c||cc||cc}
\hline
     & \multicolumn{2}{|c||}{Female} & \multicolumn{2}{|c}{Male}  \\
$M$  & loglik ($l_{M}$) & $-2(l_{M} - l_{S})$ & loglik ($l_{M}$) & $-2(l_{M} - l_{S})$ \\
\hline
\hline
0   & -40698.76  & 51.00      & -107403.60  & 42.86  \\
1   & -40690.54  & 34.56      & -107395.54  & 26.74  \\
2   & -40683.13  & 19.74*     & -107394.61  & 24.88  \\
3   & -40680.95  & 15.38      & -107393.90  & 23.46  \\
4   & -40680.69  & 14.86      & -107392.45  & 20.56  \\
5   & -40676.68  & 6.84       & -107384.24  & 4.14 *  \\
6   & -40673.26  &            & -107382.17  &   \\
\hline
\end{tabular}}
\caption{Suicide data: Log-likelihood and likelihood ratio test statistics values for the NNTS models fitted by the proposed Newton-like algorithm on the surface of a hypersphere. An asterisk (*) marks the most parsimonious models according to likelihood ratio tests using a 1$\%$ significance level.  \label{suicidesloglikAICBIC} }
\end{center}
\end{table}

\section{Conclusions}

A Newton-like algorithm on manifolds is developed to obtain the maximum likelihood estimates of the parameters of the NNTS family of distributions.
Because the parameter space corresponds to the surface of a hypersphere, other optimization methods such as sequential quadratic programming (SQP) and the
Nelder-Mead algorithm
must address norm constraints that make these optimization algorithms very slow. By working with optimization algorithms on manifolds, it is possible to avoid the use of constraints, thus making the proposed algorithm in this paper much faster and more efficient than these other methods. This is possible because the likelihood function of NNTS models can be conveniently expressed in terms of quadratic forms of the relevant parameters. The convenient use of the proposed Newton algorithm has been demonstrated in several datasets consisting
of continuous and grouped observations.

\section*{Acknowledgements}
The authors wish to thank the Asociaci\'on Mexicana de Cultura, A.C. for its support.

\end{document}